\documentclass[prb,floatfix,nobalancelastpage,twocolumn,superscriptaddress]{revtex4}
\usepackage{overpic}
\usepackage{dcolumn}% Align table columns on decimal point
\usepackage{bm}% bold math
\usepackage{amsmath, amsfonts,amssymb}
\usepackage{bm,braket}
\usepackage{color}
\usepackage{float}
\usepackage{graphicx}
\usepackage{subfigure}
\usepackage{soul}

\usepackage{sidecap}

\newcommand{\rev}[1]{{\color{red}#1}}
\newcommand{\rem}[1]{}

\definecolor{violet}{rgb}{0.58, 0.0, 0.83}

\begin{document}

\title{Floquet engineering flat bands for bosonic fractional quantum Hall with superconducting circuits}
\author{Rong-Chun Ge}
\author{Michael Kolodrubetz}
\affiliation{Department of Physics, The University of Texas at Dallas, Richardson, Texas 75080, USA}
\date{\today}

\begin{abstract}
The quest to realize novel phases of matter with topological order is an important pursuit with implications for strongly correlated physics and quantum information. Utilizing ideas from state-of-the-art coherent control of artificial quantum systems such as superconducting circuits, we present a proposal to realize bosonic fractional quantum Hall physics on small lattices by creating nearly flat topological bands using staggered flux patterns. Fingerprints of fractionalization through charge pumping can be observed with nearly perfect quantization using as few as 24 lattice sites (two photons). We suggest an implementation using a finite lattice of superconducting qubits with cylindrical connectivity on both triangular and square lattices. 

%A cold atom realization may also be possible in a one-dimensional chain of trapped atoms by exploiting internal energy levels as a synthetic dimension.
% By exploiting the idea of synthetic dimension, we show the trait of the fractional charges persists with parameter-space currently accessible in experiment.
%the phase pattern of the complex tunneling coefficient will be the multiples of half-$\pi$/one third of 2$\pi$ for the triangle and square lattices respectively; and quantum phases with fractional charges of $\nu=1/2,3/8,1/4$ are observed.
\end{abstract}
%\pacs{71.35.-y, 42.70.Qs, 71.36.+c, 85.75.-d}% PACS, the Physics and Astronomy Classification Scheme. 
\maketitle

%\mkaddcomment{
%TO DO LIST:
%\begin{enumerate}
%\item ???
%\end{enumerate}
%}
\section{Introduction}
Engineered quantum systems have galvanized the search for novel phases of matter that are not readily realized in conventional solid state systems. Topological insulators, for instance, have been extended from their original realization in electronic materials near their ground state~\cite{TI1,TI2} to distinct topological classes in Floquet~\cite{Fl1,Fl3,Fl2} and non-Hermitian systems~\cite{NonH1,NonH2,NonH3,NonH4}. Such non-traditional Hamiltonians are often demonstrated in artificial quantum systems like ultracold atoms in an optical lattice~\cite{Ultr,Ultr1} or superconducting circuits~\cite{SC,SC1}, where state-of-the-art coherent control allows terms that do not otherwise arise naturally.

The complex pattern of entanglement in topologically ordered systems makes their realization particularly challenging. The archetypal example of topological order is the fractional quantum Hall effect~\cite{FQH,FQH1,FQH2,FQH4}. Despite being realized in two-dimensional electron gases more than 30 years ago, it continues to inspire new ideas and methods to this day~\cite{FQH3,FQHn,TQC}. % \hl{[cite]}. 
An essential ingredient in quantum Hall physics is breaking time reversal symmetry. In engineered quantum systems, this has been accomplished through different schemes such as rotation~\cite{Rot,Rot1}, Raman scattering~\cite{Raman,Raman1,Raman2,Raman3,LIDG} and lattice shaking~\cite{FLC,FLC1,FLC2}, which play the role of magnetic field in conventional quantum Hall systems. 
% etc., have been employed to create synthetic gauge field in neutrally charged quantum systems, which in turn produces a vast impulse to simulate the (fractional) quantum hall physics in all kinds of available artificial quantum systems~\cite{atom,cavity,photon,three} in lab. 
Compared to the solid state, however, these artificial systems are more versatile, allowing greater addressability and control over external fields and different observables to be detected. Furthermore, engineered quantum systems can operate with fermionic or bosonic degrees of freedom, allowing avenues to bosonic fractional Hall states that are not readily available with electrons~\cite{BFQH1,BFQH2}. %\hl{[cite]}. 
Beside the fundamental questions of what phases of matter are possible, it is expected that these topologically ordered phases will have applications in quantum devices and topologically protected quantum computation~\cite{v52,FQH3}.
% Besides its importance as of fundamental research, theoretical investigations show its promise as new devices and achieving quantum computation~\cite{v52,FQH3}. Recently, the idea of synthetic dimension has been developed~\cite{synd1,synd2,synd3,synd4}, which offers new degrees of freedom and has been shown of great versatile to achieve new phases which seem to be impossible otherwise.

There has been a long theoretical history of analyzing finite size effects in fractional quantum Hall physics on small lattices~\cite{small1,small10}. These theoretic ideas have been revived in recent attempts to realize fractional quantum Hall states in engineered quantum systems~\cite{small2,small20,BFQH1}, inspired by rapidly advancing technology for fabrication and control. Recently, a state-of-the-art superconducting qubit architecture known as the ``gmon" has been developed, combining the long coherence time from the transmon qubit with fast, high fidelity control of qubit-qubit coupling \cite{SC}. Synthetic magnetic fields have been engineered in this platform by periodically modulating the coupler junctions among the qubits \cite{FLC2}. Combined with strong on-site interaction and well-developed control and measurement techniques, these advances have paved the way to achieve quantum simulations of exotic interacting many-body phases~\cite{TOPSC,MBLSC}.

In this manuscript, we present an experimental proposal to realize bosonic fractional quantum Hall physics employing gmon qubits on small two-dimensional lattices. The key insight is that, by employing a staggered flux configuration -- made possible by the novel control axes in engineered quantum systems -- a nearly flat topological Chern band can be realized. By fractionally filling this band and emulating a Laughlin-type charge pump, nearly perfect fractional quantization of topological transport may be observed for lattices as small as $6 \times 4$ sites. Furthermore, the timescales and Hamiltonian parameters used to demonstrate this are on par with those in state of the art superconducting circuits~\cite{SC}.% and ultracold atom experiments~\cite{Raman,Raman1,Raman3}.
We explicitly show that such fractional charge pumping is compatible with recent realizations of Floquet engineered fluxes in superconducting g-mon qubits \cite{SC}, which combine long coherence time with time-dependent control over hopping parameters and thus can be used to emulate hardcore bosons with complex hopping. %We also suggest how similar physics may be possible by exploiting the idea of a synthetic dimension encoded by the hyperfine energy levels such as the $|F=1/2\rangle$ manifolds of $^{87}$Rb, using one-dimensional optical lattices similar to those that realized non-interacting integer quantum Hall states~\cite{COLDQH}.
While a handful of theoretical papers have proposed to realize bosonic fractional quantum Hall physics (cf. \cite{Uterm1,Uterm2,BFQH1}), it physical realization is yet to be confirmed; we expect that within the current experimental capacity for the ``gmon" qubits by pushing the required system size down to such small values (with only 2-3 photons involved), our proposal will be an important step in the experimental endeavor to achieve the bosonic fractional quantum Hall physics in the near future. 

\section{Model and experimental configuration}
Our model Hamiltonian generalizes the periodically driven gmon system realized in \cite{FLC2}. The gmon system is based on the same planar design technology as used in Xmon, which hence inherits the long coherent time of the latter (longer than $40~\mu$s). Unlike the Xmon or other transmon-based qubits, neighboring gmons are connected by Josephson junctions, which are operated as tunable inductors. The Josephson junc-
tions enable a tunable inductive coupling between the qubits, whose amplitude can be made positive, negative, or zero. advantage of the ability to completely turn off this coupling is to avoid the frequency crowding problem~\cite{SC}, suggesting gmons as a useful platform for scaling up current small-scale quantum devices.

For any two qubits $m$ and $n$, the Hamiltonian of the system realized in experiment is~\cite{FLC2} $H_{mn}=\sum \omega_i\hat{n}_i+\sum\frac{U_i}{2}\hat{n}_i(\hat{n}_i-1)+J_{nm}(t) (a_n^\dagger a_m+h.c.)$, where $a_i^\dagger(a_i)$ are the creation (annihilation) operators of the photon mode of the $i$-th qubit, $\hat{n}_i=a_i^\dagger a_i$, $\omega_i\sim 3-5~$GHz are the frequencies of the qubits, and $U_i\sim -300~$MHz is a strong onsite nonlinear interactions of qubit excitations due to the anharmonic potential of the qubit. At leading order the system may be approximated as hardcore bosons ($U_i=\infty$).  The essence of hardcore boson limit is that the subspace without doubly occupied sites decouples effectively from the rest of the Hilbert space due to an energy mismatch. In the strong attraction limit, the high-energy manifold without doubly occupied site is well isolated, so the bosons are effectively hardcore even though one has attractive interactions, $U<0$. A time-dependent coupling $J_{mn}(t)=2J_0\cos(\Delta_{mn}t+\phi_{mn})$ ($J_0 \sim [-50,\, 5]~$MHz) is applied by modulating the external flux via the tunable inductance of the connecting junction~\cite{FLC2}. The frequency of modulation is chosen as $\Delta_{mn}=\delta\omega_{mn}$. In the rotating frame with respective to this drive, the effective Hamiltonian is
\begin{align}
 \tilde{H}_{mn} = \sum \frac{U_i}{2}a_i^\dagger a_i(a_i^\dagger a_i-1) + J_0(a_n^\dagger a_m e^{i\phi_{mn}}+h.c.),
 \end{align} 
where we have ignored the fast oscillation terms. For simplicity we will denote the Hamiltonian of the system in the rotating frame as $\tilde{H}\rightarrow H$. 
%For two qubits ($m$ and $n$) connected via coupler junctions, external time-dependent flux tuning allows control of the coupling as $J_{mn}(t)=J_0\cos(\Delta_{mn}t+\phi_{mn})$, yielding time-dependent hopping $J_{mn}(t)a^\dagger_m a_n+ h.c.$, 
%where $a^{\dagger}_{m}$ is the creation operator at site $m$. The typical value for $J_0$ is $[-50, 5]~M$Hz, which is much smaller than the characteristic drive frequencies, $\Delta_{mn}\sim $ GHz. Each qubit has an excitation energy $\omega_m \hat{n}_n$, where $\hat{n}_{m} = a^{\dagger}_{m}a_{m}$ is the number operator and $\hbar=1$ throughout. By driving on resonance, $\Delta_{mn} = \omega_n-\omega_m$, and eliminating fast oscillating terms in the rotating frame, the $\omega_{m,n}$ terms are eliminated from the Hamiltonian and the hopping term picks up a phase, $J_0 e^{i\phi_{mn}}$. 

In addition to fast control, the gmon design allows widely variable connectivity of the system~\cite{xmon}.
Furthermore, and crucially for our proposal, the phase factors on each bond can be controlled independently~\cite{FLC2}, allowing arbitrary flux patterns that are not readily realizable in conventional condensed matter systems. %Finally, due to the anharmonic potential, there are strong on-site $\hat n^2$ interactions for the qubit excitations. 
Experimentally, a one dimensional Bose-Hubbard model with the order of 10 qubits was already achieved a few years ago~\cite{OdBH1}. Recently, a finite two-dimensional (2D) lattice geometry with different size, as large as 3x7, has been shown in experiment~\cite{TdBH}. As a result, 2D lattice structures with 24-48 qubits are well reachable in principle with the current experiment.

Therefore, we start our analysi from the following Hamiltonian for gmon qubits in a two-dimensional array:
\begin{align}
H_0 = &-\sum \big(J_1^{mn}a^{\dagger}_{m+1,n}a_{m,n} + J_2^{mn}a^{\dagger}_{m,n+1}a_{m,n}\nonumber\\
+&J_3^{mn}a^{\dagger}_{m,n+1}a_{m+1,n}\big)+\frac{U}{2}\sum \hat{n}_{m,n}(\hat{n}_{m,n}-1).
\label{eq1}
\end{align}
This form is chosen to allow easy generalization to square or triangular lattices. $J_{1,2,3}^{mn} = |J_{1,2,3}^{mn}|e^{\phi_{1,2,3}^{mn}}$ are the complex nearest neighbor tunneling coefficients, with $J_3^{mn} = 0$ for square lattice, and we have neglected a uniform on-site chemical potential; $\hat{n}_{m,n} = a^{\dagger}_{m,n}a_{m,n}$ is the number operator at site $(m,n)$ with $a^{\dagger}_{m,n}$ the creation operator. For simplicity, we will first assume hardcore particles (on-site interaction $U \rightarrow \infty$ since it is generally 1 order larger than the hopping amplitude), after which we show that the relevant physics survives at finite interaction strength.  For specificity, we choose $|J_{1,2,3}^{mn}| = J= 30~$MHz, commensurate with current experimental parameters for superconducting circuits~\cite{SC}. 
Importantly, the phases of the hopping terms $J_{1,2,3}^{mn}$ will be carefully designed to give desired flux patterns as explained above. %This is readily achievable using tools of Floquet engineering, and has been accomplished in both superconductors~\cite{FLC2} and cold atoms~\cite{HarperCA1,HarperCA2}. 

\begin{figure}
\centering
\setbox1=\hbox{\includegraphics[trim=1.8cm 0.1cm 1.1cm .1cm, clip=true,height=8cm]{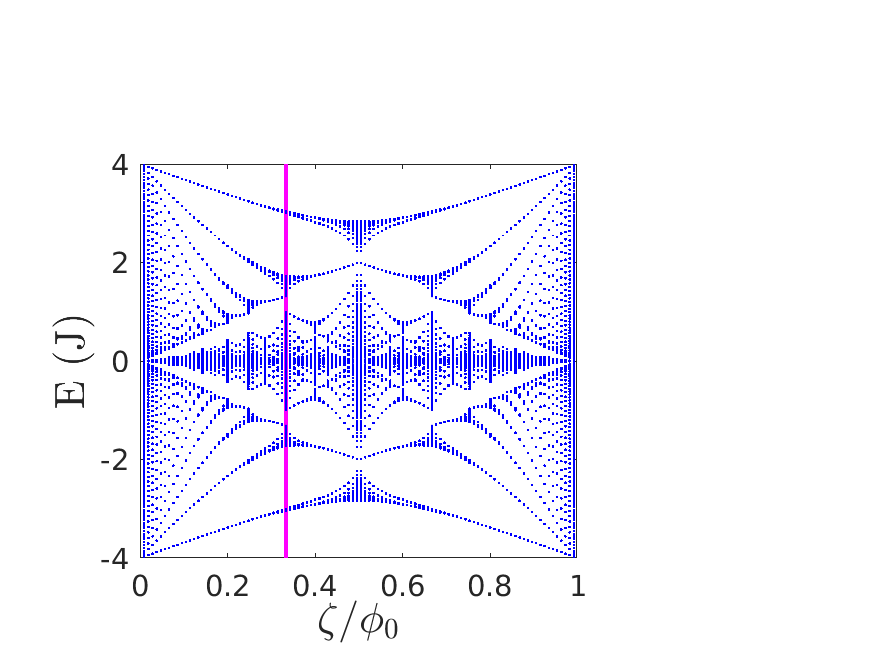}}
\includegraphics[trim=0cm 0cm 0cm 0cm, clip=true,width=0.49\columnwidth]{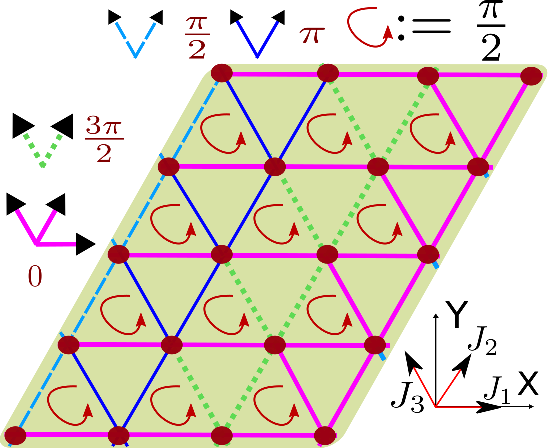}
\includegraphics[trim=0.5cm 0.2cm 4.84cm 1.0cm, clip=true,width=0.480\columnwidth]{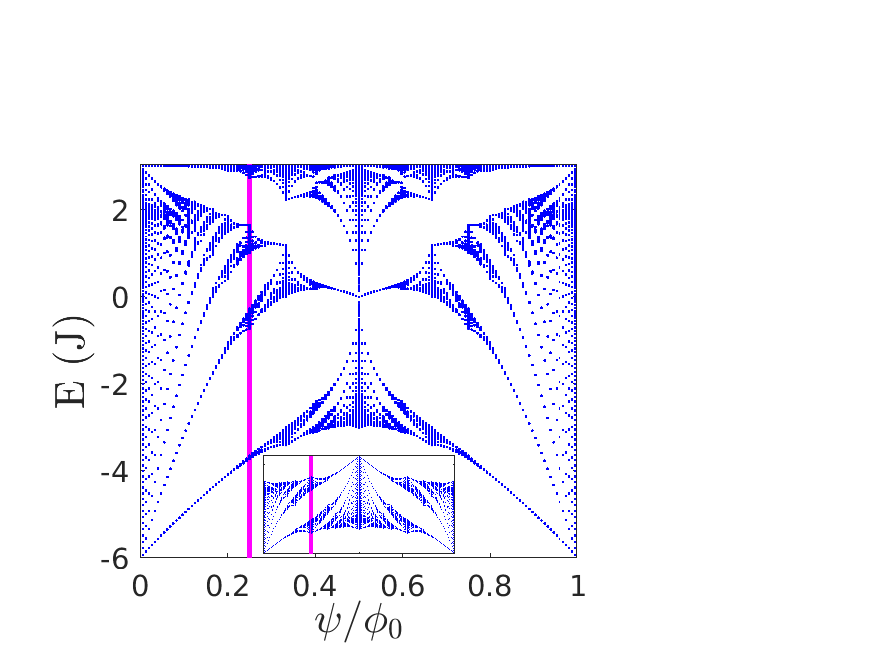}\llap{\makebox[\wd2][r]{\raisebox{0.0cm}{
\includegraphics[trim=2.9cm 15.1cm 0.1cm 12.1cm, clip=true,height=0.01cm]{newsqu2n1.eps}}}}
\includegraphics[trim=0cm 0cm 0cm .0cm, clip=true,width=0.49\columnwidth]{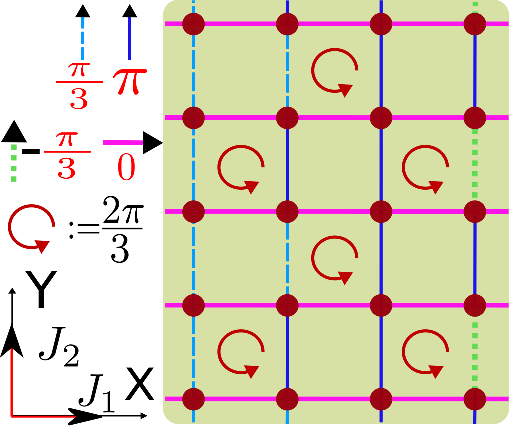}
%\begin{overpic}[trim=0.0cm 0.3cm 4.84cm 1.0cm, clip=true,width=0.499\columnwidth]{Butterflysq.eps}
\begin{overpic}[trim=0.7cm 0.2cm 4.84cm 2.0cm, clip=true,width=0.480\columnwidth]{newsqu2n1.eps}
\begin{footnotesize}
\put(0,75){{\bf\color{blue} (d)}}
\put(0,171){{\bf\color{blue} (b)}}
\put(-105,171){{\bf\color{blue} (a)}}
\put(-105,75){{\bf\color{blue} (c)}}
\end{footnotesize}
\end{overpic}
\caption{Illustration of the staggered flux lattices. (a) A triangular lattice with flux pattern $\psi/0$ in neighboring triangles shows (b) a butterfly band structure with nearly flat topological band at $\psi=\pi/2$. Coloring of bonds in (a) indicates hopping phases to realize this flux pattern. The inset to (b) shows the triangular lattice with uniform flux $\psi$ for comparison; and $\phi_0=2\pi$. (c) Staggered square lattice with flux pattern $\zeta/0$, yielding (d) topological flat band near $\zeta=2\pi/3$.}
\label{f1}
\end{figure}

\begin{figure}
\centering
\includegraphics[trim=1.2cm 0.2cm 0.4cm 1cm, clip=true,width=.99\columnwidth]{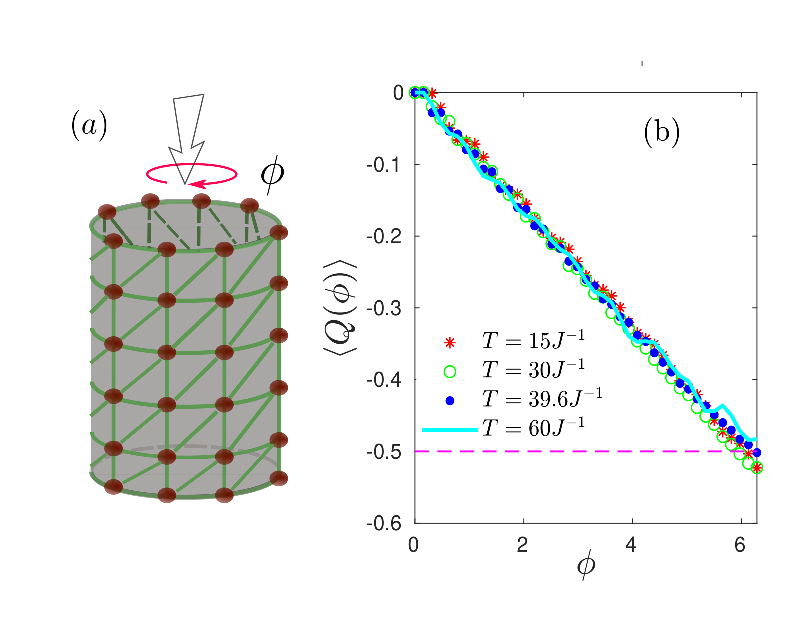}
\caption{Observation of fractional quantum Hall transport for 24 sites and filling factor $\nu=1/2$ on a triangular lattice. (a) Schematic showing flux insertion for a finite triangular lattice on a cylinder. (b) Charge transport during flux insertion as shown in (a) starting from ground state. For $J=30$ MHz, the curves correspond to $T=0.5$, $1$, $1.32$, and $2$ $\mu$s.
}
\label{f2}
\end{figure}

\section{Results and discussion}
{\it Triangular lattice --} We first consider a triangular lattice with flux configuration shown in Fig.~\ref{f1}(a), with flux $\psi$ and $0$ in adjacent triangles respectively. Experimentally, this can be achieved by having $J_1^{mn} = J$, $J_2^{mn} = J\exp\left[im\psi\right]$ and $J_3^{mn} =J\exp\left[i(m+1)\psi\right]$.
In order to optimize the working parameters to achieve the fractional quantum Hall states, we first obtain the band structure in the absence of interactions (analogous to the Hofstadter butterfly~\cite{Hbutter}) as a function of flux $\psi$. The results, shown in Fig.~\ref{f1}(b), indicate significant band flattening compared to uniform flux (Fig.~\ref{f1}(b), inset). %In particular, at $\psi = \pi/2$, there are four bands, with the bottom band nearly flat;
We are most interested in the bottom band, since it will dominate the many-body (interacting) ground state. At $\psi = \pi/2$, we consider a figure of merit for band flatness give by the ratio of the gap, $\Delta$, to the bandwidth, $W$~\cite{small2}: $M \equiv \frac{\Delta}{W} \approx 33.84$. This large $M$ indicates that the band is very flat. For $\psi=2\pi/3$ as  chosen in~\cite{BFQH1}, $M\approx 7.29$; as a result, we will work with $\psi=\pi/2$.

We first consider the lattice on a torus with twisted boundary conditions to understand the topology of the states. We calculated the many-body Chern number~\cite{MBChern,ChernCal},
\begin{align}
C = \frac{1}{2\pi}\int d\phi_1 d\phi_2 {\text Tr}\left[\frac{\partial A^2}{\partial \phi_1}-\frac{\partial A^1}{\partial \phi_2} \right]
\end{align}
where $A^\alpha_{mn} = -i\langle m_{\phi_1,\phi_2}|\partial_{\phi_\alpha}|n_{\phi_1,\phi_2}\rangle$ is the Berry connection matrix calculated within the degerate ground state sector. At half filling, $\nu = 1/2$, for a $6\times 4$ lattice we have $3$ particles, and the ground state manifold is two dimensional. The calculation gives $C=1=\frac12+\frac12$ for the ground state manifold, consistent with a fractional $\nu=1/2$ bosonic quantum Hall state~\cite{torus}. Further calculations confirm that this topological band is robust for most other system sizes at half filling, and for the non-interacting case we find $C=1$ in the thermodynamic limit.

Experimentally, the hallmark of the fractional quantum Hall effect is conductance in a Hall bar geometry, but this is not practical in such small, isolated quantum systems. Instead, we consider the theoretically proposed Laughlin charge pump~\cite{gauge1,gauge2}, 
which we argue is actually realizable in engineered settings. When a quantum Hall state (at thermodynamic limit) is prepared on a ring or cylinder and a single quantum of magnetic flux is adiabatically inserted through the center, an electric field will induced around the cross of the cylinder (see below), and some effective (fractionally) charge will be pumped from one side to the other,  which gives the quantized Hall conductivity. In this case, the quantized charge ($Q_{2\pi}$) is expected to be equal to the filling factor $\nu$~\cite{FQH2}. We notice that a similar drift idea has been exploited to detect the fractional quantum states in cold atomic system~\cite{Detect} recently.

We test quantized transport by placing the system on a cylinder with periodic boundary conditions in the $J_2$ direction and open boundary conditions in the $J_1$ direction, as shown in Fig.~\ref{f1}(a). The system is initially prepared in the ground state when no flux ($\phi=0$) is injected through the cylinder. Then the flux is ramped up uniformly at a finite speed by modulating the phase across the circumference of the cylinder. The charge transported as a function of the injected flux  is obtained as
\begin{align}
%\langle Q(\phi)\rangle = \frac{P(\phi)-P(\phi=0)}{L},
\langle Q_{\phi}\rangle = \frac{1}{L}\int_0^\phi \partial_{\phi^\prime}P(\phi^\prime)d\phi^\prime,
\end{align}      
where $L$ is the length of cylinder in the open ($J_1$) (open boundary) direction and $P(\phi)=\langle \sum_{i}x_i\hat{n}_i\rangle_{\phi}$ is the many-body polarization in $J_1$ direction, with $\langle \cdots \rangle_{\phi}$ the average over the evolving state of the system at flux $\phi$. In experiment, at the end of the flux pumping, $\phi=2\pi$, all the coupling between the qubits can be turned off, which makes it easy to measure the charge distribution.
%For a finite system with $M$ particles, the polarization can be obtained explicitly as $P(\phi)=\langle \sum_{m=1}^M (i_m-1)n_{i_m,i'_m}\rangle_{\phi}$. \mkaddcomment{\hl{Is this the best way to write this, in first-quantized notation?}}\rep{Is this better?}
%\mkaddcomment{``For $m<l$ we require $i_m\leq i_l$ and $(i_m,i'_m)\neq (i_l,i'_l)$; if $i_m=i_l$ we futher require $i'_m<i'_l$ for $m<l$'' -- MK: I don't understand this statement.}\rep{We can just delete this, I was trying to describe how to do this calculation numerically.}
%bare bosons/particles and the lattice constant is 1;  the information of the number of particles $\langle \sum_{m=1}^M n_{i_m,i'_m}\rangle_{\phi}=M$ has been included implicitly in the wavefunction since the number of particles is a good quantum number. 
%Here $\langle \cdots \rangle_{\phi}$ is the average over the state at flux $\phi$. 
We consider flux which is injected at a constant rate $\dot \phi = 2 \pi / T$, which is equivalent to adding a small transverse electric field $E_y = -\dot \phi / L_y$.

Our numerical calculations show that for a lattice as small as 24 sites, the $\nu=1/2$ fractional pumping survives with a quantization accurate to 99.6\%, as shown by the blue dots in Fig.~\ref{f2}(b). Importantly, this data corresponds to finite pumping time $T \in 15-60~J^{-1}$, rather than adiabatic flux insertion ($T \to \infty$). %This is because there is energy splitting between the degenerate ground states due to finite size effects;
The relevant physics can be understood from the many-body band structure shown in Fig.~\ref{sf1}(b) in the Appendix. Due to finite-size effects, a non-vanishing coupling between the topological ground state and an excited state is observed at certain non-zero flux, as indicated by the circled anti-crossing. Consequently, in order to drive the system along what would be the adiabatic path in the thermodynamic limit, a finite speed is required to cross these finite size gaps diabatically. % topological pumping requires crossing this gap diabatically.
Numerically, we have confirmed that the results are not significantly affected by changing $T$ within about 50\% of the optimal value, $T_\mathrm{opt} \sim 40/J$, which is estimated to be maximally diabatic with respect to the ground state manifold and adiabatic with respect to excitations out of this manifold. The observed quantized charge pumping breaks down when the pumping time $T$ deviates significantly from this optimal value, as can be seen in the supplement in the Appendix, Fig.~\ref{sf1}(a).

{\it Square lattice --} While triangular lattices have certain advantages, square lattices are often more natural experimentally. Therefore, we attempt to find the same physics on square lattices, employing the flux configuration shown in Fig.~\ref{f1}(c). The single particle energy spectrum (Fig.~\ref{f1}(d)) shows flat, low energy bands near $\zeta = 2\pi/3$ with $M\approx27.88$. 
The first two bands are well separated from the higher energy bands, and we find that they together yield $C=1$ suggesting that they are amenable to fractional Hall physics at partial filling. 

\begin{figure}
\centering
\includegraphics[trim=0.3cm 4.2cm 0.8cm 0.0cm, clip=true,width=.99\columnwidth]{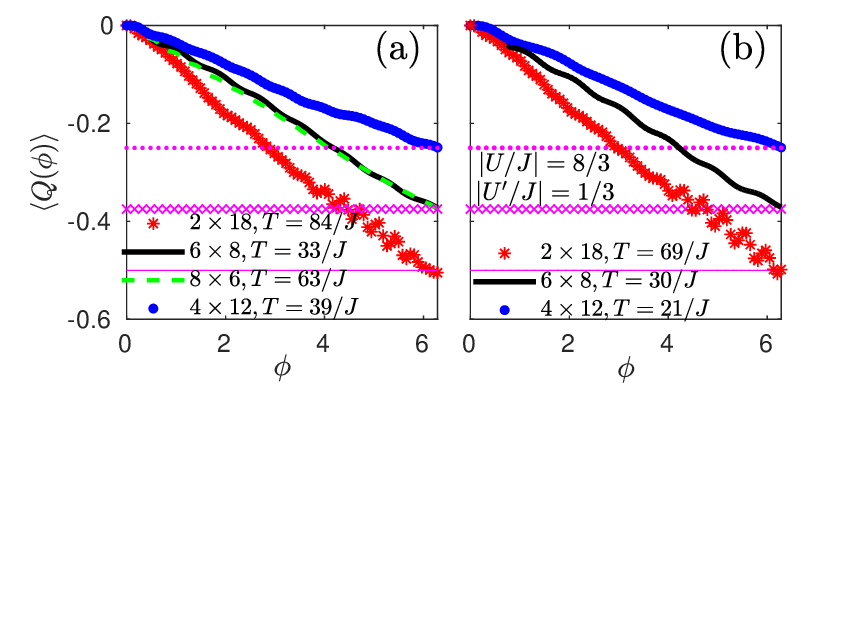}
\caption{Realizing fractional quantum Hall physics with a finite square lattice using g-mon qubits in with cylindrical connectivity.  (a) Fractional charge pumping for hardcore bosons on a square lattice ($U'=0$). (b) Fractional charge pumping for non-hardcore bosons with finite interactions between sites in the $y$-direction (see Eq.~\ref{eq:H_Ad}). %\hl{Remove Figure 3c, leave it for later work.}
}
\label{f4}
\end{figure}

Our numerical simulations confirm this prediction. For $\nu=1/2$, the fractional Hall effect survives to system size as small as 2 lattice sites in the $y$ (periodic) direction and 18 sites in $x$ direction (Fig.~\ref{f4}(a), red stars).  A well defined quantum charge pump, ($Q_{2\pi}=0.5$ indicated by the magneta solid line) could be observed for a relatively large time window ($[0.5,\,\,\, 3.5]$~$\mu$s for $J=30~M$Hz), and the result shown is for $T=84~J^{-1}$ which is $2.8~\mu$s for $J=30~M$Hz. 
%The result shown here is the average over the time window $[0.5,\,\,\, 4.4]$~$\mu$s. 
The evolution of the energy as a function of $\phi$ shows a similar minimal gap (similar $T_c$) as for the triangular lattice. %\rev{Calculations of the charge distribution, $\rho_x=\sum_y\langle \hat{n}_{x,y}\rangle$, for the ground state as shown in Fig.~\ref{f4}(c) with zero injected flux $\phi=0$ confirm that ground state at weak injected flux corresponds to the edge state of interest; this is consistent with the recent results using density matrix renormalization group technique~\cite{DMRGfinite} in which a uniform flux distribution was employed.} %Plots of the local charge density confirm that the lower energy states correspond to the edge states~\cite{SOM}. %\hl{***put this in the supplement and cite it if you have time***}

By changing the system size, we can achieve other fractional quantum Hall states on the square lattice as well. For $L_x \times L_y = 6 \times 8$ or $8 \times 6$ and filling $\nu=3/8$, a robust $Q_{2\pi}=0.375$ (shown by the magenta crosses) pumping is observed (Fig.~\ref{f4}(a)). This state may be understood with the composite fermion picture~\cite{FQH2,BFQH1} in the continuum limit, in which three vortices are binded to a boson with $n=3$ fully filled Landau levels, $\nu = \frac{n}{3n-1}$. 
%particle is being transported; all these analysis indicates that the $Q_{2\pi}=3/8$ factional quantum hall states could be accessible experimentally for a system of 48 sites. 
The solid black in  shows the result for size $6\times8$ with $T~1.1\mu$s ($|J|=30~M$Hz); the green dashed is a single calculation for size $8\times6$.
Other fractional fillings give well-quantized transport, such as $\nu=1/2$ for $2\times 18$, $\nu=3/8$ for $6\times 8$ or $8\times 6$, and $\nu=1/4$ for $4\times 12$, as shown in Fig.~\ref{f4}(a). For fractions and lattice sizes that did not give strong fractional Hall signals, a key issue is lack of separation between the ``small'' gap between topological and non-topological states due to finite size effects and the ``large'' gap to other excited states, preventing the existence of a wide time window for quantized pumping. These gaps are due to complex and difficult to unravel finite size effects, which has prevented us from obtaining any further insight on the exact conditions for quantized pumping to occur.

%{\it Synthetic dimension --}  
With ``gmons'', one would directly create the lattices and complex hoppings by doing time-dependent shaking of the potentials, as realized experimentally in \cite{FLC2}. The qubit excitations are known to be well-approximated by tunneling hardcore bosons and have long coherence time ($T_1$ is around a few tens to a few hundreds of micro-seconds, much larger than time window used to observe the quantized pumping). The system sizes proposed of order 20-40 qubits are commensurate with state-of-the-art experiments~\cite{IBM20,SCreview}.

%\rep{{\it Synthetic dimension --} 
We point out that the same idea can be applied to ultracold atoms in an optical lattice. A synthetic dimension consisting of four atomic hyperfine levels could be employed to create the cylindrical geometry. Synthetic dimensions have been successfully used to demonstrate non-interacting topological phases, including many of the key ingredients of our proposal such as using artificial gauge fields -- created by a pair of Raman beams -- to break time-reversal symmetry~\cite{Raman,Raman1,Raman2}. Adding and controlling these particular flux patterns require additional control fields. A detailed proposal and analysis is beyond the scope of this paper, but will be presented in forthcoming work. 

To demonstrate stability of our proposed topological response, here we consider a simplified model of the cold atom system on a %\hl{[cite Spielman]}. 
%As an explicit example, we consider realizing the 
$4\times 12$ square lattice.
%using the 5$^2S_{1/2}$, $F=1$ and $F=2$ levels of $^{87}$Rb. We propose to use an external magnetic field along the $z$-direction to split the $m_F$ levels, %\mkaddcomment{Does it matter what we consider x/y/z here?}, 
%after which laser beams could be employed as in Fig.~\ref{f4}(c) to the $|F=1,m_F=0/\pm1\rangle$ and $|F=2,m_F=0\rangle$ state. 
A length-4 cycle is formed in the synthetic dimension (y), with hopping phases controlled to give flux patterns as described above. 
%by spin dependent tunneling phase. 
Unlike real dimensions, synthetic dimensions have interspecies interactions between atoms on the same physical site. We model this effect here by adding isotropic interactions between the distinct hyperfine levels:
\begin{align}
H_{Ad} = U'\sum_{n\neq n'} \hat{n}_{m,n}\hat{n}_{m,n'}.
\label{eq:H_Ad}
\end{align}
Figure~\ref{f4}(b) shows that for finite on site Hubbard interaction and weak adjacent interaction, $U/J = 8/3,U'/J=1/3$, the near-perfect quantized pumping remains.
% with the current experiment setup of the superconducting qubits (for which $U'=0$ and is believed to favor the fractional states) 
%And the above adjacent interaction, $H_{Ad}$ will describe interaction among the spins at the same lattice site. For $|J_{1/2}|=3$~kHz , and $|U/J_{1/2}|=8$, $|U'/J|=1/3$, the typical ti%me need to obtain the quantized pumping is a few tens of ms.
%It can be shown that previous observed quantized pumping persists for even a relative weak interaction strength. More generally we introduce a adjacent interaction besides the on-site interaction Hubbard interaction:

\section{Conclusion}
In summary, we have presented an experimental proposal to observe the bosonic fractional quantum Hall effect physics employing superconducting circuits on both triangular and square lattices with a relative small lattice size. Only nearest neighboring tunneling is required in our proposal, and the phases required for the nearest neighbor tunneling are within existing experimental capability. %\mkaddcomment{Should we add a few words about time scales of Floquet drive and why we can just use the effective Hamiltonian picture?}
%multiples of $\pi/2$ for triangular lattice and multiples of $2\pi/3$ for square lattice, which makes them more accessible with the experiment technique at hands. 
For g-mon qubits, the non-trivial tunneling phase is obtained by periodically modulating the coupling strength $J_i^0$. Since the frequency of the modulation ($\omega$) is much larger than the coupling strength $|J_i^0|$, at the short or medium time limits considered here, the dynamics is well approximated by the effective Hamiltonian used in this work.
Similar experiments are possible by exploiting synthetic dimensions made from hyperfine levels of $^{87}$Rb in a one-dimensional optical lattice.

\begin{figure}
\centering
\includegraphics[trim=0.0cm 0.0cm 0.2cm 0.0cm, clip=true,width=.99\columnwidth]{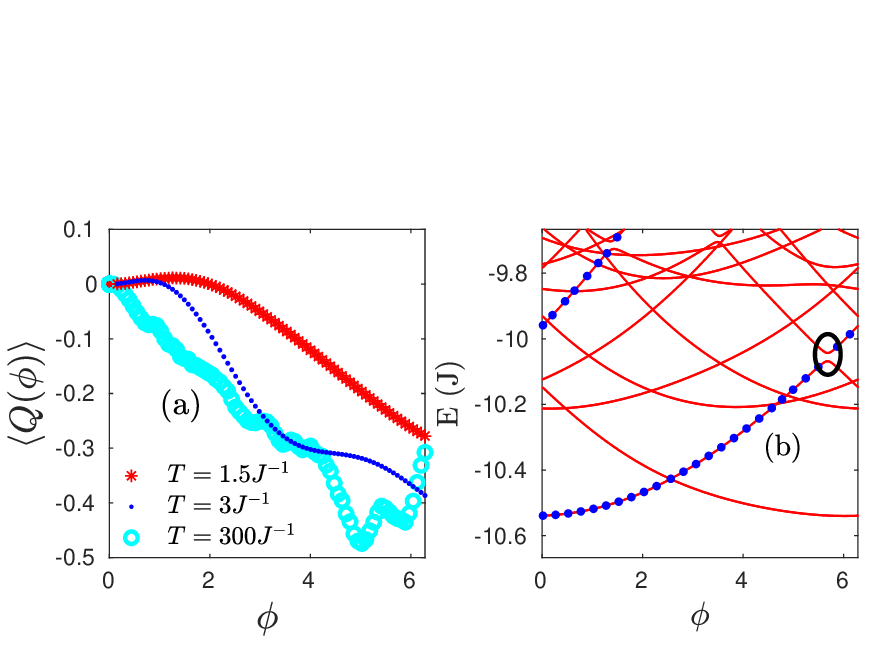}
\caption{ 24 sites triangular lattice on cylinder with half filling $\nu = 1/2$ (hardcore bosons).  (a) Deviation from the fractional quantized charge transport ($Q_{2\pi}=1/2$) when the flux pumping period $T$ is too small/large compared to the optimum value as shown in the main text. (b) Many-body band-structure of the lattice on the cylinder as a function of the flux, $\phi$, along the axis of cylinder. The blue dots indicated the desired ground state that we want to "adiabatically" track. The black circle shows the finite size gap which must be crossed diabatically. The shortest time scale for doing so is consistent with the optimum $T$..}
\label{sf1}
\end{figure}

This would be the first experimental demonstration of bosonic fractional Hall physics and of fractional Chern insulators, bosonic or fermionic, in engineered systems. We note that the flat bands demonstrated here by Floquet engineering are not special to bosonic systems, so we expect that similar topologically ordered phases should be achievable in fermionic systems as well.
\section{Acknowledgments}
We would like to acknowledge useful discussions with P. Roushan. This work was performed with support from the National Science Foundation through award number DMR-1945529 and the Welch Foundation through award number AT-2036-20200401. We used the computational resources of the Lonestar 5 cluster operated by the Texas Advanced Computing Center at the University of Texas at Austin and the Ganymede and Topo clusters operated by the University of Texas at Dallas' Cyberinfrastructure \& Research Services Department

\appendix
%\addcontentsline{toc}{section}{Appendices}
\section*{Appendix}
\label{secap}

Figure.~\ref{sf1}(a) shows the deviation from the quantized charge transport ($Q_{2\pi}$) for the 24-site triangular lattice at half-filling when the flux-pumping time is too small ($T=0.05~\mu$s (red stars), $T=0.1~\mu$s (blue dots) for $J=30~$MHz) or too large ($T=10~\mu$s) compared to optimum time which is around $T=1.32~\mu$s. For very fast ramping speed the system will be excited out of its ground state manifold,  which breaks the quantized transport. 

For these relatively small finite size systems, a large time also causes issues because one then adiabatically crosses non-topological gaps that are opened due to finite size effects. This can be seen in the many body level diagram of Fig. 4(b).

Following the general wisdom of adiabatic pumping, for initial state prepared as the ground state at $\phi=0$, we will need to follow the path illustrated by the blue dots; the first relevant anti-crossing is indicated by the black circle. The time scale required to cross this finite size gap diabatically is consistent with the optimum time $T=1.32~\mu$s found numerically for $|J|=30~$MHz.

%\begin{align}
%H = \sum \big(J_1^{mn}a^{\dagger}_{m+1,n}a_{m,n} + J_2^{mn}a^{\dagger}_{m,n+1}a_{m,n}\big)+\frac{U}{2}\sum \hat{n}_{m,n}(\hat{n}_{m,n}-1) + U'\sum_{n\neq n'} \hat{n}_{m,n}\hat{n}_{m,n'}
%\end{align}

%%%%%%%%%%%%%%%%%%%%%%% References %%%%%%%%%%%%%%%%%%%%%%%%%


\begin{thebibliography}{[1]}
\bibitem{TI1}
M. Z. Hasan, and C. L. Kane, Rev. Mod. Phys. {\bf 82}, 3045(2010)
\bibitem{TI2}
X.-L. Qi, and S.-C. Zhang, Rev. Mod. Phys. {\bf 83}, 1057 (2011)
%\bibitem{TI3}
%C. L. Kane, and E. J. Mele, Phys. Rev. Lett. {\bf 95}, 226801 (2005)
%\bibitem{TI4}
%C. L. Kane, and E. J. Mele, Phys. Rev. Lett. {\bf 95}, 146802 (2005)
\bibitem{Fl1}
T. Kitagawa, E. Berg, M. Rudner, and E. Demler, Phys.Rev. B {\bf 82}, 23114 (2010)
\bibitem{Fl3}
M. C. Rechtsman, J. M. Zeuner, Y. Plotnik, Y. Lumer, D. Podolsky, F. Dreisow, S. Nolte, M. Segev, and A. Szameit Nature {\bf 496}, 196 (2013) 
\bibitem{Fl2}
M. S. Rudner, N. H. Lindner, E. Berg, and M. Levin, Phys. Rev. X {\bf 3}, 031005 (2013)
\bibitem{NonH1}
T. E. Lee, Phys. Rev. Lett. {\bf 116}, 133903 (2016)
\bibitem{NonH2}
H. Shen, B. Zhen, and L. Fu, Phys. Rev. Lett. {\bf 120}, 146402 (2018)
\bibitem{NonH3}
S. Yao and Z. Wang, Phys. Rev. Lett. {\bf 121}, 086803 (2018)
\bibitem{NonH4}
Z. Gong, Y. Ashida, K. Kawabata, K. Takasan, S. Higashikawa, and M. Ueda, Phys. Rev. X {\bf 8}, 031079 (2018)
\bibitem{Ultr}
I. Bloch, J. Dalibard, and W. Zwerger, Rev. Mod. Phys. {\bf 80}, 885 (2008)
\bibitem{Ultr1}
C. Chin, R. Grimm, P. Julienne, and E. Tiesinga, Rev. Mod. Phys. {\bf 82}, 1225 (2010)
\bibitem{SC}
Y. Chen, C. Neill, P. Roushan, N. Leung, M. Fang, R. Barends, J. Kelly, B. Campbell, Z. Chen, B. Chiaro, A. Dunsworth, E. Jeffrey, A. Megrant, J. Y. Mutus, P. J. J. O'Malley, C. M. Quintana, D. Sank, A. Vainsencher, J. Wenner, T. C. White, M. R. Geller, A. N. Cleland, and J. M. Martinis, Phys. Rev. Lett. {\bf 113}, 220502 (2014)
\bibitem{SC1}
Z. Yan, Y.-R. Zhang, M. Gong, Y. Wu, Y. Zheng, S. Li, C. Wang, F. Liang, J. Lin, Y. Xu1, C. Guo, L. Sun, C.-Z. Peng, K. Xia, H. Deng, H. Rong, J. Q. You, F. Nori, H. Fan, X. Zhu, and J.-W. Pan, Science {\bf 364}, 753 (2019)
\bibitem{FQH}
D. C. Tsui, H. L. Stormer, and A. C. Gossard, Phys. Rev. Lett. {\bf 48}, 1559 (1982)
\bibitem{FQH1}
R. B. Laughlin, Phys. Rev. Lett. {\bf 50}, 1395 (1983)
\bibitem{FQH2}
J. K. Jain, Phys. Rev. Lett. {\bf 63}, 199 (1989)
\bibitem{FQH4}
M. Kaicher, S. B. J\"{a}ger, P.-L. Dallaire-Demers, and F. Wilhelm,	Phys. Rev. A {\bf 102} 022607 (2020)
\bibitem{TQC}
A. Y. Kitaev, Ann. Phys. {\bf 303}, 2 (2003)
\bibitem{FQH3}
C. Nayak, S. H. Simon, A. Stern, M. Freedman, and S. D. Sarma, Rev. Mod. Phys. {\bf 80}, 1083 (2008)
\bibitem{FQHn}
T. H. Hansson, M. Hermanns, S. H. Simon, and S. F. Viefers, Rev. Mod. Phys. {\bf 89}, 025005 (2017) 
\bibitem{Rot1}
A. L. Fetter, Rev. Mod. Phys. {\bf 81}, 647 (2009)
\bibitem{Rot}
N. R. Cooper, Adv. Phys. {\bf 57}, 539 (2008)
\bibitem{Raman}
Y.-J. Lin, R. L. Compton, A. R. Perry, W. D. Phillips, J. V. Porto, and I. B. Spielman, Phys. Rev. Lett. {\bf 102}, 130401 (2009)
\bibitem{Raman1}
Y.-J. Lin, R. L. Compton, K. Jim\'{e}nez-García, J. V. Porto, and I. B. Spielman, Nature {\bf 462}, 628 (2009).
\bibitem{Raman2}
A. Celi, P. Massignan, J. Ruseckas, N. Goldman, I. B. Spielman, G. Juzeliunas, and M. Lewenstein, Phys. Rev. Lett. {\bf 112}, 043001 (2014)
\bibitem{Raman3}
J. Dalibard, F. Gerbier, G. Juzeliunas, and P. \"{O}hberg, Rev. Mod. Phys. {\bf 83}, 1523 (2011)
\bibitem{LIDG}
N. Goldman, G Juzeliunas, P \"{O}hberg and I B Spielman, Rep. Prog. Phys. {\bf 77}, 126401 (2014)
\bibitem{FLC}
P. Hauke, O. Tieleman, A. Celi, C. \"{O}lschl\"{a}ger, J. Simonet, J. Struck, M. Weinberg, P. Windpassinger, K. Sengstock, M. Lewenstein, and A. Eckardt, Phys. Rev. Lett. {\bf 109}, 145301 (2012)
\bibitem{FLC1}
N. Goldman, and J. Dalibard, Phys. Rev. X {\bf 4} 031027 (2014)
\bibitem{FLC2}
P. Roushan, C. Neill, A. Megrant, Y. Chen, R. Babbush, R. Barends, B. Campbell, Z. Chen, B. Chiaro, A. Dunsworth, A. Fowler, E. Jeffrey, J. Kelly, E. Lucero, J. Mutus, P. J.J. O'Malley, M. Neeley, C. Quintana, D. Sank, A. Vainsencher, J. Wenner, T. White, E. Kapit, H. Neven, and J. Martinis, Nat. Phys. {\bf 13}, 146 (2017)
\bibitem{BFQH1}
N. R. Cooper, and J. Dalibard, Phys. Rev. Lett. {\bf 110}, 185301 (2013)
\bibitem{BFQH2}
N. R. Cooper, J. Dalibard, I. B. Spielman, Rev. Mod. Phys. {\bf 91}, 015005 (2019)
\bibitem{v52}
G. Moore, and N. Read, 1991, Nucl. Phys. B {\bf 360}, 362 (1991)
\bibitem{small1}
G. S. Kliros, and N d'Ambrumenil, J. Phys. Condens. Matter {\bf 3} 4241 (1991)
\bibitem{small10}
G. S. Kliros, Physica A {\bf 183}, 209 (1992)
\bibitem{small2}
S. Kourtis, J. W. F. Venderbos, and M. Daghofer, Phys. Rev. B {\bf 86}, 235118 (2012)
%Fractional Chern insulator on a triangular lattice of strongly correlated  t2g electrons
\bibitem{small20}
J. W. F. Venderbos, M. Daghofer, and J. van den Brink, Phys. Rev. Lett. {\bf 107}, 116401 (2011) 
%\bibitem{small3}
%N. R. Cooper and J. Dalibard, Phys. Rev. Lett. {\bf 110}, 185301 (2013)
\bibitem{TOPSC}
P. Roushan, C. Neill, Yu Chen, M. Kolodrubetz, C. Quintana, N. Leung, M. Fang, R. Barends, B. Campbell, Z. Chen, B. Chiaro, A. Dunsworth, E. Jeffrey, J. Kelly, A. Megrant, J. Mutus, P. J. J. O’Malley, D. Sank, A. Vainsencher, J. Wenner, T. White, A. Polkovnikov, A. N. Cleland, and J. M. Martinis, Nature, {\bf 515}, 241 (2014)
\bibitem{MBLSC}
C. Neill, P. Roushan, M. Fang, Y. Chen, M. Kolodrubetz, Z. Chen, A. Megrant, R. Barends,
B. Campbell, B. Chiaro, A. Dunsworth, E. Jeffrey, J. Kelly, J. Mutus, P. J. J. O’Malley, C. Quintana, D. Sank, A. Vainsencher, J. Wenner, T. C. White, A. Polkovnikov, and J. M. Martinis, Nat. Phys. {\bf 12} 1037 (2016).

%\bibitem{COLDQH}
%B. K. Stuhl, H.-I. Lu, L. M. Aycock, D. Genkina, I. B. Spielman, Science {\bf 349}, 1514 (2015)
\bibitem{Uterm1}
Y.-F. Wang, Z.-C. Gu, C.-D. Gong, and D. N. Sheng, Phys. Rev. Lett. {\bf 107}, 146803 (2011)

\bibitem{Uterm2}
P. Rosson, M. Lubasch, M. Kiffner, and D. Jaksch, Phys. Rev. A {\bf 99}, 033603 (2019)
\bibitem{xmon}
R. Barends, J. Kelly, A. Megrant, D. Sank, E. Jeffrey, Y. Chen, Y. Yin, B. Chiaro, J. Mutus, C. Neill, P. O'Malley, P. Roushan, J. Wenner, T. C. White, A. N. Cleland, and J. M. Martinis, Phys. Rev. Lett. {\bf 111}, 080502 (2013)
\bibitem{OdBH1}
P. Roushan, C. Neill, J. Tangpanitanon, V.M. Bastidas, A. Megrant, R. Barends, Y. Chen, Z. Chen, B. Chiaro, A. Dunsworth, A. Fowler, B. Foxen, M. Giustina, E. Jeffrey, J. Kelly, E. Lucero, J. Mutus, M. Neeley, C. Quintana, D. Sank, A. Vainsencher, J. Wenner, T. White, H. Neven, D. G. Angelakis, J. Martinis, Science {\bf 358}, 1175 (2017)
\bibitem{TdBH}
B. Chiaro, {\it et al.,} arXiv:1910.06024
%\bibitem{FQE1}
%M. Bukov, L. D'Alessio, and A. Polkovnikov, Adv. Phys. {\bf 64}, 139 (2015).
%\bibitem{FQE2}
%P. Weinberga, M. Bukova, L. D’Alessioa, A. Polkovnikova, S. Vajnaa, and M. Kolodrubetz, Phys. Rep. {\bf 688}, 1 (2017).


%\bibitem{BOSEH}
%M. Greiner, O. Mandel, T. Esslinger, T. W H\"{a}nsch, and I. Bloch, Nature {\bf 415}, 39 (2002)
%\bibitem{HarperCA1}
%M. Aidelsburger, M. Atala, M. Lohse, J. T. Barreiro, B. Paredes, and I. Bloch, Phys. Rev. Lett. {\bf 111}, 185301 (2013) 
%\bibitem{HarperCA2}
%H. Miyake, G. A. Siviloglou, C. J. Kennedy, W. C. Burton, and W. Ketterle, Phys. Rev. Lett. {\bf 111}, %185302 (2013)
\bibitem{Hbutter}
D. R. Hofstadter, Phys. Rev. B {\bf 14}, 2239 (1976)
\bibitem{MBChern}
Q. Niu, D. J. Thouless, and Y. Wu
Phys. Rev. B {\bf 31}, 3372 (1985)
\bibitem{ChernCal}
T. Fukui, Y. Hatsugai, and H. Suzuki, J. Phys. Soc. Jpn. {\bf 74}, 1674 (2005)
\bibitem{torus}
X.G. Wen, Int. J. Mod. Phys. B {\bf 4}, 239 (1990)
\bibitem{gauge1}
R. B. Laughlin, Phys. Rev. B {\bf 23}, 5632 (1981)
\bibitem{gauge2}
B. I. Halperin, Phys. Rev. B {\bf 25}, 2185 (1982)
\bibitem{Detect}
C. Repellin, J. L\'{e}onard, and N. Goldman, Phys. Rev. A {\bf 102}, 063316 (2020) 
%\bibitem{DMRGfinite}
%P. Rosson, M. Lubasch, M. Kiffner, and D. Jaksch, Phys. Rev. A {\bf 99}, 033603 (2019)
\bibitem{IBM20}
 G. J. Mooney, C. D. Hill, and L. C. L. Hollenberg, Sci. Rep {\bf 9},13465(2019).
\bibitem{SCreview}
M. Kjaergaard, M. E. Schwartz, J. Braum\"{u}ller, P. Krantz, J. I-J Wang, S. Gustavsson, and W. D.Oliver, Annu. Rev. Condens. Matter Phys. {\bf 11}, 369 (2020) 
\rev{
%\bibitem{Qsupre}
%F. Arute, {\it et al.,} Nature {\bf 574}, 505 (2019)
%\bibitem{prepare}
%The details about the preparation of the initial state is beyond the regime of our current work, and is left for future work. 
}
%\bibitem{atom}
%A. S. S{\o}rensen, E. Demler, and M. D. Lukin, Phys. Rev. Lett. {\bf 94}, 086803 (2005)
%\bibitem{cavity}
%J. Cho, D. G. Angelakis, and S. Bose, Phys. Rev. Lett. {\bf 101}, 246809 (2008)
%\bibitem{photon}
%R. O. Umucalilar and I. Carusotto, Phys. Rev. Lett. {\bf 108}, 206809 (2012)
%\bibitem{three}
%E. Kapit, and S. H. Simon, Phys. Rev. B {\bf 88}, 184409 (2013)
%\bibitem{synd1}
%E. Lustig, S. Weimann, Y. Plotnik, Y. Lumer, M. A. Bandres, A. Szameit, and M. Segev, Nature {\bf 567}, 356 (2019)
%\bibitem{synd2}
%T. Ozawa, and H. M. Price,  Nat. Rev. Phys. {\bf 1}, 349 (2019)
%\bibitem{synd3}
%G. Salerno, H. M. Price, M. Lebrat, S. H\"{a}usler, T. Esslinger, L. Corman, J.-P. Brantut, and N. Goldman, Phys. Rev. X {\bf 9}, 041001 (2019)
%\bibitem{synd4}
%H. M. Price, T. Ozawa, and N. Goldman, Phys. Rev. A {\bf 95}, 023607 (2017)
%\bibitem{finite0}
%A. Kol, and N. Read, Phys. Rev. B {\bf 48}, 8890 (1993)
%\bibitem{finite1}
%M. Hafezi, A. S. S{\o}rensen, E. Demler, and M. D. Lukin, Phys. Rev. A {\bf 76}, 023613 (2007)
%\bibitem{finite2}
%F. M\"{o}ller, and N. R. Cooper, Phys. Rev. Lett. {\bf 103}, 105303 (2009)
%\bibitem{QSatom}
%C. Gross, and I. Bloch, Science {\bf 357}, 995 (2017) 
%\bibitem{synd1}
%H. M. Price, T. Ozawa, and N. Goldman, 
%\bibitem{synd2}
%T. Ozawa, and H. M. Price,  Nat. Rev. Phys. {\bf 1}, 349 (2019) 
%\bibitem{syndouble}
%A. Celi, P. Massignan, J. Ruseckas, N. Goldman, I.B. Spielman, G. Juzeliunas, and M. Lewenstein, Phys. Rev. %Lett. {\bf 112}, 043001 (2014)




%\bibitem{Wen}




 

%\end{thebibliography} 
\end{thebibliography}
\end{document}